\documentclass[12pt]{article}
\begin{document}
\begin{center}
{\Large {\bf A hidden symmetry in the Standard Model}}

\vskip-40mm \rightline{\small ITEP-LAT/2003-32} \vskip 30mm

{
\vspace{1cm}
{B.L.G.~Bakker$^a$, A.I.~Veselov$^b$, M.A.~Zubkov$^b$  }\\
\vspace{.5cm} { \it $^a$ Department of Physics and Astronomy,
Vrije Universiteit, Amsterdam,
The Netherlands \\
$^b$ ITEP, B.Cheremushkinskaya 25, Moscow, 117259, Russia }}
\end{center}

\begin{abstract}
We found an additional symmetry hidden in the fermion and Higgs sectors of the
Standard Model. It is connected to the centers of the $SU(3)$ and $SU(2)$
subgroups of the gauge group. A lattice regularization of the whole Standard
Model is constructed that possesses this symmetry.
\end{abstract}


\section{Introduction}

It is well known that to put a quantum field theory onto a lattice one
should keep as much symmetries of the original model as possible. That
is why, for example, any lattice gauge model is made to preserve the
gauge symmetry \cite{creutz} while it is possible, in principle,  to
construct a lattice model that comes as a discretization of a gauge
fixed continuum theory. Other examples of this kind are the attempts to
put fermions on a lattice both avoiding  doubling and keeping the
chiral symmetry \cite{lattice_fermions}.

It is the conventional point of view that all the symmetries of
the Standard Model (SM), which must be used when dealing with its
discretization, are known.  In this letter we demonstrate (in the
framework of lattice regularization) that an additional symmetry
is hidden within the fermion and Higgs sectors of the SM. It is
connected to the centers of the $SU(3)$ and $SU(2)$ subgroups. It
turns out possible to redefine the gauge sector of the lattice
realization of the SM in such a way that it has the same naive
continuum limit as the conventional one, while keeping the
additional symmetry.

The Standard Model contains the following variables:

1. The gauge field ${\cal U} = (\Gamma, U, \theta)$, where
\begin{eqnarray}
 \Gamma \in SU(3), \quad U \in SU(2), \quad e^{i\theta} \in U(1),
\end{eqnarray}
realized as link variables on the lattice.

2. A scalar doublet
\begin{equation}
 \Phi^{\alpha}, \;\alpha = 1,2.
\end{equation}

3. Anticommuting spinor variables, representing leptons and
quarks:
\begin{equation}
 \left(
 \begin{array}{ccc}
  \nu_e & \nu_{\mu} &  \nu_{\tau}\\
  e & \mu & \tau ,
 \end{array}
 \right) , \quad
 \left(
 \begin{array}{ccc}
 u & c & t \\
 d & s & b
 \end{array}
 \right) .
\end{equation}

The action has the form
\begin{equation}
 S = S_g + S_H + S_f,
\end{equation}
where we denote the fermion part of the action by $S_{f}$, the
pure gauge part is denoted by $S_g$, and the scalar part of the
action by $S_H$.

In {\it any} lattice realization of $S_H$ and $S_f$ both these terms
depend upon link variables $\cal U$ considered in the representations
corresponding to quarks, leptons, and the Higgs scalar field,
respectively. Therefore $\cal U$ appears in the combinations shown in
the table.
\begin{table}
\label{tab.01}
\begin{center}
\begin{tabular}{|c|l|}
\hline
$U\, e^{-i\theta}$ & {\rm left-handed leptons} \\
\hline
$e^{-2 i \theta}$ & {\rm right-handed leptons} \\
\hline
$ \Gamma \, U \, e^{ \frac{i}{3} \theta}$ & {\rm left-handed quarks}\\
\hline
$ \Gamma \, e^{ -\frac{2i}{3} \theta}$ &{\rm right-handed d, s, and, b - quarks} \\
\hline
$ \Gamma \, e^{ \frac{4i}{3} \theta}$ &{\rm right-handed u, c, and, t - quarks} \\
\hline
$  U \, e^{  i \theta}$ &{\rm the Higgs scalar field}\\
\hline
\end{tabular}
\end{center}
\end{table}
Our observation is that {\it all} the listed combinations are
invariant under the following transformations:
\begin{eqnarray}
 U & \rightarrow & U e^{-i\pi N}, \nonumber\\
 \theta & \rightarrow & \theta +  \pi N, \nonumber\\
 \Gamma & \rightarrow & \Gamma e^{(2\pi i/3)N},
\label{symlat}
\end{eqnarray}
where $N$ is an arbitrary integer link variable. It represents a
three-dimensional hypersurface on the dual lattice. Both $S_H$ and
$S_f$ (in {\it any} realization) are invariant under the simultaneous
transformations (\ref{symlat}). This symmetry reveals the
correspondence between the centers of the $SU(2)$ and $SU(3)$
subgroups of the gauge group.

After integrating out fermion and scalar degrees of freedom any
physical variable should depend upon gauge invariant quantities
only. Those are the Wilson loops: $\omega_{SU(3)}({\cal C}) = {\rm
Tr} \Pi_{{\rm link} \in {\cal C}} \Gamma_{\rm link}$,
$\omega_{SU(2)}({\cal C}) = {\rm Tr} \Pi_{{\rm link} \in {\cal C}}
U_{\rm link}$, and $\omega_{U(1)}({\cal C}) = \Pi_{{\rm link} \in
{\cal C}} {\rm exp}(\frac{i}{3} \theta_{\rm link})$ Here $\cal C$
is an arbitrary closed contour on the lattice (with self -
intersections allowed). These Wilson loops are trivially
invariant under the transformation (\ref{symlat}) with the field
$N$ representing a {\it closed} three-dimensional hypersurface on
the dual lattice. Therefore the nontrivial part of the symmetry
(\ref{symlat}) corresponds to a closed two-dimensional surface
on the dual lattice that is the boundary of the hypersurface
represented by $N$. Then in terms of the gauge invariant quantities
$\omega$ the transformation (\ref{symlat}) acquires the form:
\begin{eqnarray}
 \omega_{U(1)}({\cal C}) & \rightarrow &
 {\rm exp}(-i \mbox{\small $\frac{1}{3}$} \pi {\bf L} ({\cal C}, \Sigma))
 \, \omega_{U(1)}({\cal C}) \nonumber\\
 \omega_{SU(2)}({\cal C}) &
 \rightarrow & {\rm exp}( i \pi {\bf L} ({\cal C}, \Sigma))
 \, \omega_{SU(2)}({\cal C}) \nonumber\\
 \omega_{SU(3)}({\cal C}) & \rightarrow &
 {\rm exp}( i \mbox{\small $\frac{2}{3}$} \pi {\bf L}
 ({\cal C}, \Sigma))\, \omega_{SU(3)}({\cal C}) \label{sym}
\end{eqnarray}
Here $\Sigma$ is an arbitrary closed surface (on the dual lattice)
and ${\bf L} ({\cal C}, \Sigma)$ is the integer linking number of this
surface and the closed contour $\cal C$.

It is worth mentioning that after integrating out fermion degrees
of freedom as well as the Higgs scalar the Standard Model in its
continuum formulation becomes a theory defined in a loop space
\cite{loop_equations}, i.e. any physical variable depends upon
gauge fields only through the $SU(3), SU(2)$ an $U(1)$ Wilson
loops. If we again denote them as $\omega_{SU(3)}$,
$\omega_{SU(2)}$, and $\omega_{U(1)}$ (where $\omega_{U(1)}$
corresponds to the worldline of a particle of $U(1)$ charge
$\frac{1}{3}$ while $\omega_{SU(2)}$ and $\omega_{SU(3)}$ are the
Wilson loops considered in the fundamental representations of $SU(2)$
and $SU(3)$ respectively), the symmetry (\ref{sym}) understood in
the continuum notation would appear if we neglect the pure
gauge-field part of the action. It is obvious that the latter in
its conventional continuum formulation (or, say, in lattice Wilson
formulation) is not invariant under (\ref{sym}). However, the
lattice realization of the pure gauge field term of the action can
be constructed in such a way that it also preserves the mentioned
symmetry. For example, we can consider the following expression
for $S_g$:
\begin{eqnarray}
 S_g & = & \sum_{\rm plaquettes}
 \{\beta_1(1-\mbox{${\small \frac{1}{2}}$} {\rm Tr}\, U_p \cos \theta_p)
 \nonumber \\
 && + \beta_2 (1-\cos 2\theta_p) \nonumber \\
 && + \beta_3 (1-\mbox{${\small \frac{1}{6}}$} {\rm Re Tr}
 \,\Gamma_p {\rm Tr}\, U_p \exp (i\theta_p/3))
 \nonumber\\
 && +\beta_4(1-\mbox{${\small \frac{1}{3}}$} {\rm Re Tr}
 \, \Gamma_p \exp (-2i\theta_p/3)) \nonumber \\
 && +\beta_5(1-\mbox{${\small \frac{1}{3}}$} {\rm Re Tr}
 \, \Gamma_p \exp (4i\theta_p/3))\},\label{Act}
\end{eqnarray}
where the sum runs over the elementary plaquettes of the lattice.
Each term of the action Eq.~(\ref{Act}) corresponds to a parallel
transporter along the boundary of a plaquette considered in one of
the representations listed above. The coefficients $\beta_i, \, (i
= 1, \dots 5)$ must be chosen in such a way as to give rise to the
correct value of the Weinberg angle.

Naively Eq.~(\ref{Act}) has the same continuum limit (with the
appropriate choice of $\beta_i$) as, say the following conventional
action:
\begin{eqnarray}
 S^0_g & = & \sum_{\rm plaquettes}
 \{\beta^0_1(1-\mbox{${\small \frac{1}{2}}$} {\rm Tr}\, U_p )
 \nonumber \\
 && + \beta^0_2 (1-\cos \theta_p) \nonumber \\
 && + \beta^0_3 (1-\mbox{${\small \frac{1}{3}}$} {\rm Re Tr}
 \,\Gamma_p )\},\label{Act0}
\end{eqnarray}
However, (\ref{Act}) possesses the additional symmetry (\ref{sym}) while
(\ref{Act0}) does not. If the symmetry (\ref{sym}) does occur in nature, a
regularization that does not maintain it would be inappropriate. The situation
here could be similar to that of an attempt to construct a lattice gauge model
while not keeping the gauge invariance: the corresponding lattice model may
describe physics improperly.

A particularly interesting question is how the symmetry (\ref{symlat})
emerges in lattice discretizations of unified models.  Namely,
(\ref{symlat}) may naturally appear after the breakdown $G \rightarrow
SU(3) \otimes SU(2) \otimes U(1)$. The simplest example of the unified
model of such type is the conventional $SU(5)$ theory \cite{okun}. If
we consider its lattice definition with the Wilson action, the low
energy limit would coincide with Eq.~(\ref{Act}) for the following choice
of couplings:
\begin{equation}
\beta_{1} = 2\beta/5,\quad \beta_{4} = 3\beta/5,\quad \beta_2 = \beta_3 =
\beta_5 = 0.
\end{equation}
Relation (\ref{symlat}) itself appears to be the trivial consequence of
expressing $SU(5)$ link matrices in terms of $\Gamma, U$ and $\theta$ in the
low energy approximation:
\begin{equation}
\left(\begin{array}{cc}
 \Gamma e^{-\frac{2i\theta}{3}} & 0 \\
 0 & U e^{i \theta}
 \end{array} \right)
\end{equation}
The same picture emerges in any unified theory, if its gauge group $G$
contains $SU(5)$ and the symmetry breakdown pattern is $G \rightarrow
\dots \rightarrow SU(5) \rightarrow SU(3)\otimes SU(2) \otimes U(1)$.

The other unified models may be transferred to the lattice either
violating or preserving (\ref{symlat}). As an example, let us consider
the $SU(2)_L\otimes SU(2)_R\otimes SU(4)'_{L+R}$ Pati - Salam unified
model \cite{PS}. We arrange the fermions (of the first generation)
as the elements of $2\times 4$ matrices ${\cal F}^{ab}_{L,R}$ (the
$SU(2)_{L,R}$ subgroups act on the first index, the $SU(4)'$ subgroup
acts on the second index):
\begin{equation}
{\cal F}_{L,R} = \left(\begin{array}{cccc}
 u^1 & u^2 & u^3 & \nu \\
 d'^1 & d'^2 & d'^3 & e
 \end{array} \right)_{L,R}.
\label{F}
\end{equation}
Leptons and quarks of the other generations are arranged in a similar way.

Let us construct the Higgs sector in such a way that it provides link
matrices which have the form (at low energies, after the breakdown
$SU(2)_L\otimes SU(2)_R\otimes SU(4) \rightarrow SU(3) \otimes SU(2)_L
\otimes U(1)$):
\begin{equation}\begin{array}{ccccc}
 U &\otimes &
 \left(\begin{array}{cc}
  e^{ i \theta}& 0 \\
  0 & e^{-i \theta }
 \end{array} \right)&\otimes &
 \left(\begin{array}{cc}
 \Gamma e^{\frac{i\theta}{3}} & 0 \\
 0 &  e^{- i \theta}
 \end{array} \right).
 \end{array}
\end{equation}

We can define the pure gauge field action, say, in the following two ways:

1. Let ${\cal V} = {\cal Y}^L\otimes {\cal Y}^R\otimes {\cal Z} \in
SU(2)_L\otimes SU(2)_R \otimes SU(4)$ be the $SU(2)_L\otimes SU(2)_R\otimes
SU(4)$ link matrix (here ${\cal Y}^{L,R}\in SU(2), {\cal Z} \in SU(4)$). Then
let us consider the action of the form:
\begin{equation}
 S = \beta \sum_{\rm plaq}\{(1 - \frac{1}{2}{\rm Re}\,{\rm Tr}\,
 {\cal Y}^L_{\rm plaq}) + (1 - \frac{1}{2}{\rm Re}\,{\rm Tr}\,
 {\cal Y}^R_{\rm plaq}) + (1 - \frac{1}{4}{\rm Re}\,{\rm Tr}\,
 {\cal Z}_{\rm plaq})\}
\end{equation}
The lattice model defined in this way obviously violates (\ref{symlat}) in the
low energy limit.

2. With the above definition of the link variable let us now consider the
lattice model with another action
\begin{eqnarray}
 S & = & \beta \sum_{\rm plaq}(1 - \frac{1}{16}{\rm Re}\,({\rm Tr}\,
 {\cal Y}^L_{\rm plaq} + {\rm Tr}\,
 {\cal Y}^R_{\rm plaq})\,{\rm Tr}\, {\cal Z}_{\rm plaq})\nonumber\\
 & \sim & \beta \sum_{\rm plaq}(1 - \frac{1}{16}{\rm Re}\,({\rm Tr}\,
 U_{\rm plaq} + 2 \cos(\theta_{\rm plaq}))({\rm Tr} \Gamma_{\rm plaq}
 e^{\frac{i}{3} \theta_{\rm plaq}} + e^{-i \theta_{\rm plaq}}))
\nonumber \\
\label{act4}
\end{eqnarray}
This is exactly the action (\ref{Act}) with the following choice of
couplings:
\begin{equation}
\beta_{1} =  \beta/8,\quad
\beta_{2} = \beta/16,\quad
\beta_3 = 3 \beta/8,\quad
\beta_{4} = \beta_5 = 3\beta/16.
\label{couplings}
\end{equation}
Therefore, the full unified model preserves our additional symmetry after
the breakdown $SU(2)_L\otimes SU(2)_R \otimes SU(4) \rightarrow SU(3)
\otimes SU(2)_L \otimes U(1)$.

Finally, we consider a unified model with {\it arbitrary} gauge group $G$
and the arrangement of fermions such that there exist representations
$\alpha, \gamma, ... $ of $G$ that are {\it completely} composed of the
full set of Standard Model fermions. Let again ${\cal V} \in G$ be the
link variable. We choose the action \begin{equation} S = \beta_{\alpha}
\sum_{\rm plaq}(1 - {\rm Re}\,\chi_{\alpha} ({\cal V}_{\rm plaq}))
+ \beta_{\gamma} \sum_{\rm plaq}(1 - {\rm Re}\,\chi_{\gamma} ({\cal
V}_{\rm plaq}))+...,\label{act0} \end{equation} where $\chi_{\alpha}$
is the character of the representation $\alpha$ and the sum is over the
mentioned representations. The resulting model preserves (\ref{symlat})
after the breakdown $G \rightarrow SU(3)\otimes SU(2) \otimes U(1)$. We
like to mention here that Eq.~(\ref{Act}) with the couplings given by
Eq.~(\ref{couplings}) (the $SU(2)\otimes SU(2)\otimes SU(4)$ model)
would appear also in the low energy limit of the $SU(5)$ unified model
if the action of the latter is chosen as the sum of (\ref{act0}) -
like terms corresponding to both representations, in which the fermions
are arranged. This happens because in both cases the action (\ref{act0})
involves all the representations that exhaust the full set of the Standard
Model fermions.

So, the symmetry (\ref{symlat}) being confirmed (or rejected) would
give a criterium for the choice of a unified model. The dynamical
consequence of (\ref{symlat}) could appear due to the fact, that it ties
the centers of the $SU(3)$ and $SU(2)$ subgroups of the gauge group. It
is well-known that the center elements of the color subgroup of the gauge
group play an important role in the description of the confinement of
color \cite{BVZ00,BVZ2002,Greensite1,Greensite}. Therefore one might
expect that in the model with the pure gauge field action (\ref{Act})
it may not be possible to investigate color dynamics alone (without
taking into account the $SU(2)$ or $U(1)$ subgroups of the gauge group)
and the confinement picture may be different from the one found within
the framework of the conventional discretization.

On the other hand, the topological excitations corresponding to
the center of the $SU(2)$ subgroup may play an important role in the
finite temperature nonperturbative electroweak phenomena
\cite{EW_T}. Therefore due to the mentioned ties, the
description of, say, the finite temperature electroweak phase
transition may also be different for the lattice models which do
or do not maintain the additional symmetry.

A comparison of the two approaches in these respects may be
important for understanding whether it is necessary or not to take
into account the additional symmetry considered, while
constructing the lattice approximation to the Standard Model.

\vspace{5ex}

We are grateful to M.I. Polikarpov, F.V. Gubarev and V.A. Rubakov
for useful discussions. A.I.V. and M.A.Z. kindly acknowledge the
hospitality of the Department of Physics and Astronomy of the
Vrije Universiteit, where part of this work was done. This work
was partly supported by RFBR grants 01-02-17456, 03-02-16941 and
02-02-17308, by the INTAS grant 00-00111, the CRDF award
RP1-2364-MO-02, DFG grant 436 RUS 113/739/0 and RFBR-DFG grant
03-02-04016, by Federal Program of the Russian Ministry of
Industry, Science and Technology No 40.052.1.1.1112.


\begin{thebibliography}{99}

\bibitem{creutz}
M.~Creutz,
{\em Quarks, gluons and lattices}, (Cambridge University Press,
Cambridge, 1985).

\bibitem{lattice_fermions}
N.B.~Nielsen and M.~Ninomiya,
Nucl. Phys. {\bf  B185}, 20 (1981); {\it ibid}, 173;

M.~L{\"{u}}scher, Phys. Lett. {\bf B428}, 342 (1998);

H.~Neuberger, Phys. Lett. {\bf B417}, 141 (1998)

\bibitem{loop_equations}
 Y.M.~Makeenko and A.A.~Migdal, Nucl.Phys.
{\bf B188}, 269 (1981);

A.A.~Migdal, Phys.Rep. {\bf 102}, 199 (1983).

\bibitem{okun}
L.B.~Okun {\em Leptons and quarks}, (North Holland, Amsterdam, 1982).

\bibitem{PS} J.C.Pati, S.Radjpoot, A.Salam, Phys. Rev.  {\bf D 17}, 131 (1978)

\bibitem{BVZ00}
B.L.G. Bakker, A.I. Veselov, and M.A. Zubkov,\\
Phys. Lett. {\bf B 502} (2001) 338; hep-lat/0011062

\bibitem{BVZ2002}
B.L.G.~Bakker, A.I.~Veselov, and M.A.~Zubkov Phys.Lett. {\bf
B544}, 374 (2002);

\bibitem{Greensite1}
M. Faber, J. Greensite, and S. Olejnik, in ``Confinement,
Topology, and other Non-Perturbative Aspects of QCD'', NATO
Advanced Research Workshop, Stara Lesna, Slovakia, 2002

\bibitem{Greensite}
J. Greensite,
Prog. Part. Nucl. Phys. 51 (2003) 1

\bibitem{EW_T}
M. Gurtler, E.M. Ilgenfritz, and A. Schiller,
Phys. Rev. {\bf D56}, 3888 (1997);

M.N. Chernodub, F.V. Gubarev, E.M. Ilgenfritz, and A. Schiller
Phys. Lett. {\bf B443}, 244 (1998).

\end{thebibliography}
\end{document}